\documentclass{mem}
\usepackage{natbib}\usepackage{txfonts}\usepackage{balance}
\usepackage{graphicx}
\usepackage{ulem}
\usepackage[breaklinks, colorlinks, citecolor=blue, urlcolor = blue]{hyperref}
\idline{75}{282}

\begin{document}
\def\th{THESEUS}
\def\aone{$\alpha_1$}
\def\atwo{$\alpha_2$}
\def\eb{$E_{\rm break}$}
\def\ep{$E_{\rm peak}$}
\def\be{$\beta$}

\title{
THESEUS and Gamma-Ray Bursts
}
\subtitle{a valuable contribution to the understanding of prompt emission}

\author{
Lara \,Nava\inst{1,2,3} 
\and Gor \,Oganesyan\inst{4}
\and Maria~E. \,Ravasio\inst{1,5}
\and Lorenzo Amati\inst{6}
\and \\Giancarlo Ghirlanda\inst{1,5,7}
\and Paul \,O'Brien\inst{8}
\and Julian~P.\,Osborne\inst{8}
\and Richard \,Willingale\inst{8}
          }

\institute{
INAF --
Osservatorio Astronomico di Brera, Via Bianchi 46,
I-23807 Merate, Italy\\
\email{lara.nava@brera.inaf.it}
\and
INAF --
Osservatorio Astronomico di Trieste, Via Tiepolo 11,
I-34131 Trieste, Italy
\and
INFN --
Sezione di Trieste, via Valerio 2, 
I-34127 Trieste, Italy
\and
SISSA, via Bonomea 265,
I-34136 Trieste, Italy
\and
Universit\`a degli Studi di Milano-Bicocca, 
Dipartimento di Fisica U2, 
Piazza della Scienza, 
3, I-20126, Milano, Italy
\and
INAF -- IASF Bologna, 
via P. Gobetti, 101, 
I-40129 Bologna, Italy 
\and
INFN –- Sezione di Milano-Bicocca, Piazza della Scienza 3, I-20126 Milano, Italy
\and
Department of Physics and Astronomy, 
University of Leicester, 
Leicester LE1 7RH, United Kingdom
}

\authorrunning{Nava L.}

\titlerunning{The THESEUS contribution to GRB prompt emission}

\abstract{
Recent advances in fitting prompt emission spectra in gamma-ray bursts (GRBs) are boosting our understanding of the still elusive origin of this radiation.
These progresses have been possible thanks to a more detailed analysis of the low-energy part ($<$\,100\,keV) of the prompt spectrum, where the spectral shape is sometimes  found to deviate from a simple power-law shape.
This deviation is well described by a spectral break or, alternatively by the addition of a thermal component. 
Spectral data extending down to less than 1\,keV are extremely relevant for these studies, but presently they are available only for a small subsample of {\it Swift} GRBs observed by XRT (the X-ray telescope, 0.3-10\,keV) during the prompt emission. 
The space mission \th\ will allow a systematic study of prompt spectra from 0.3\,keV to several MeV. 
We show that observations performed by \th\ will allow us to discriminate between different models presently considered for GRB prompt studies, solving the long-standing open issue about the nature of the prompt radiation, with relevant consequences on the location of the emitting region, magnetic field strength and presence of thermal components.

\keywords{ }
}
\maketitle{}

\section{Introduction}\label{sec:introduction}
Spectral information on prompt emission from gamma-ray bursts (GRBs) became available for a large sample ($\sim$\,2700 events) and on a wide range of energies ($\sim$\,25\,keV -- 2\,MeV) first thanks to the Burst And Transient Source Experiment (BATSE), the soft $\gamma$-ray instrument onboard the {\it Compton Gamma Ray Observatory}, in orbit from 1991 to 2000.
Since the very first studies, prompt spectra revealed their non-thermal nature, requiring a (partial) dissipation of the outflow energy and the consequent energisation of a non-thermal population of accelerated particles.
The most natural candidate for the radiation process appeared to be leptonic synchrotron emission.
However, it was soon realised that for a sizeable fraction of GRBs, the low-energy part of the observed spectrum is harder than (and then inconsistent with) synchrotron predictions \citep{preece98}. 

For reasonable properties of the emitting region, the electrons are expected to radiate in a regime of fast cooling \citep{ghisellini00}.
This would produce a synchrotron photon spectrum with power-law index $\alpha_2^{\rm syn}=-1.5$ (in the notation $N_{\rm E}\propto E^{\alpha_2}$) below the $\nu F_{\nu}$ spectral peak $\nu_{\rm peak}$, extending unperturbed at lower frequencies (if self-absorption is negligible), until reaching the so-called cooling break $\nu_{\rm c}$, with $\nu_{\rm c}\ll\nu_{\rm peak}$.
Below $\nu_{\rm c}$ the synchrotron spectrum is still described by a power-law (PL), but with a harder photon index ($\alpha_1^{\rm syn}=-2/3$).
For most GRBs, observations rule out such interpretation in terms of fast cooling synchrotron spectra: below the spectral peak, prompt spectra are well described by one single PL segment, with typical photon index $\langle\alpha\rangle\simeq-1$. 

A regime of {\it marginally fast} cooling has been proposed as a plausible explanation for the apparent inconsistency: in this regime $\nu_{\rm c}$ and $\nu_{\rm peak}$ are close to each other ($\nu_{\rm c}\lesssim \nu_{\rm peak}$, implying a still large radiation efficiency) and the asymptotic index at low energies $\nu<[\nu_{\rm c},\nu_{\rm peak}]$ quickly approaches the value $\alpha_1^{\rm syn}=-2/3$ \citep{daigne11}.  
This scenario can in principle explain photon indices up to -2/3, but does not solve the problem of spectra harder than this limiting synchrotron value, for which, in the context of synchrotron models, self-absorption must be invoked. 

A completely different approach to the problem consists in turning to thermal models.
Besides a few examples where the observed spectral shape is consistent with a pure thermal spectrum (e.g. \citealt{ghirlanda03}), in the vast majority thermal models face the opposite problem as compared to synchrotron models: a thermal spectrum is too hard to explain observations.
However, various reasonable processes can easily soften the resulting spectra, such as the convolution of multi-temperature black-bodies (BB), or the composition of an underlying non-thermal spectrum and a thermal component.

The simultaneous presence of a BB-like component and a non-thermal component has been tentatively identified in a several cases (still a relatively small fraction of GRBs).
These two-component models can be classified into two different cases: i) the spectral peak is dominated by the thermal component (see e.g. \citealt{ryde10}), and ii) the spectral peak is dominated by the non-thermal component, while the thermal emission contributes to the flux only at lower energies (see e.g. \citealt{guiriec16}).

Recently, a major advancement in the characterisation of prompt spectra has been reported. 
Thanks to those sporadic cases (34 studied so far) where the prompt emission has been (at least partially) observed also by the X-ray telescope (XRT, 0.3-10\,keV) onboard {\it The Neil Gehrels Swift Observatory} (hereafter {\it Swift}), the frequent presence of a spectral break located between 2\,keV and 20\,keV was discovered \citep{oganesyan17,oganesyan18}.
Below the break energy, the spectrum is well described by a PL function. 
A similar result has been recently found also in GRB~160525B \citep{ravasio18}, one of the brightest GRBs ever detected by the {\it Fermi}-GBM (8\,keV-1\,MeV). 
For this GRB, XRT observations during the prompt emission were not available, but the break energy is located around $\sim$\,100\,keV, well within the GBM range of sensitivity. 
GBM data alone were sufficient to constrain the energy of this spectral break and the index of the PL below the break, with no need for soft X-ray observations.

Remarkably, the spectral analysis of these spectra featuring a break in the low-energy part, revealed a general agreement with the synchrotron model: the spectral slope below and above the break are, on average, consistent with the synchrotron slopes if the break identified at a few keV corresponds to the cooling frequency, and the peak energy \ep\ corresponds to the characteristic synchrotron energy. 

For some of these GRBs, previous spectral studies proposed a different modeling:
the convolution of a thermal and a non-thermal components.
The two different models (single component with low-energy break and two-components) fit the spectra equally well above a few keV.
However, they predict very different behaviours at lower energies. 
A larger number of spectra with data extending to energies $<1$\,keV and covering the full prompt emission phase is required before reaching a conclusion of the viability of the two different models. 

The study of this energy domain for a larger sample of GRBs and the consequent progresses in the characterisation of prompt spectra are within the reach of the \th\ mission \citep{amati17, stratta17}.
Thanks to the joint effort of the SXI (Soft X-ray Imager, 0.3-6\,keV) and XGIS (X-Gamma rays Imaging Spectrometer, 2\,keV-20\,MeV) instruments, \th\ will ensure a simultaneous coverage on a broad energy range (from 0.3\,keV to 20\,MeV combining the SXI and XGIS) from the trigger time of the detected GRBs.

\begin{figure*}[t!]
\resizebox{\hsize}{!}
{\includegraphics[clip=true,scale=0.57]{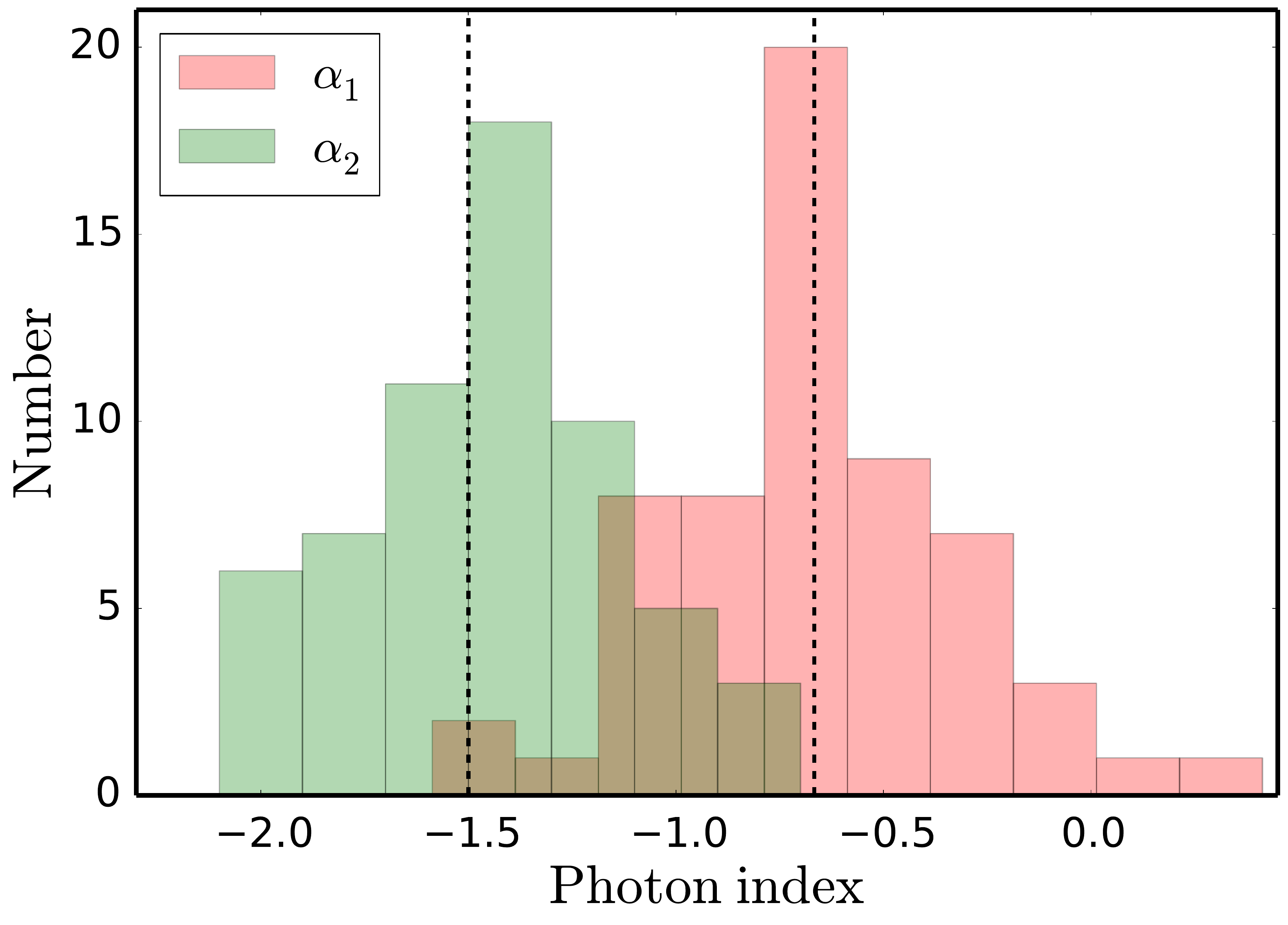}
\includegraphics[clip=true,scale=0.87]{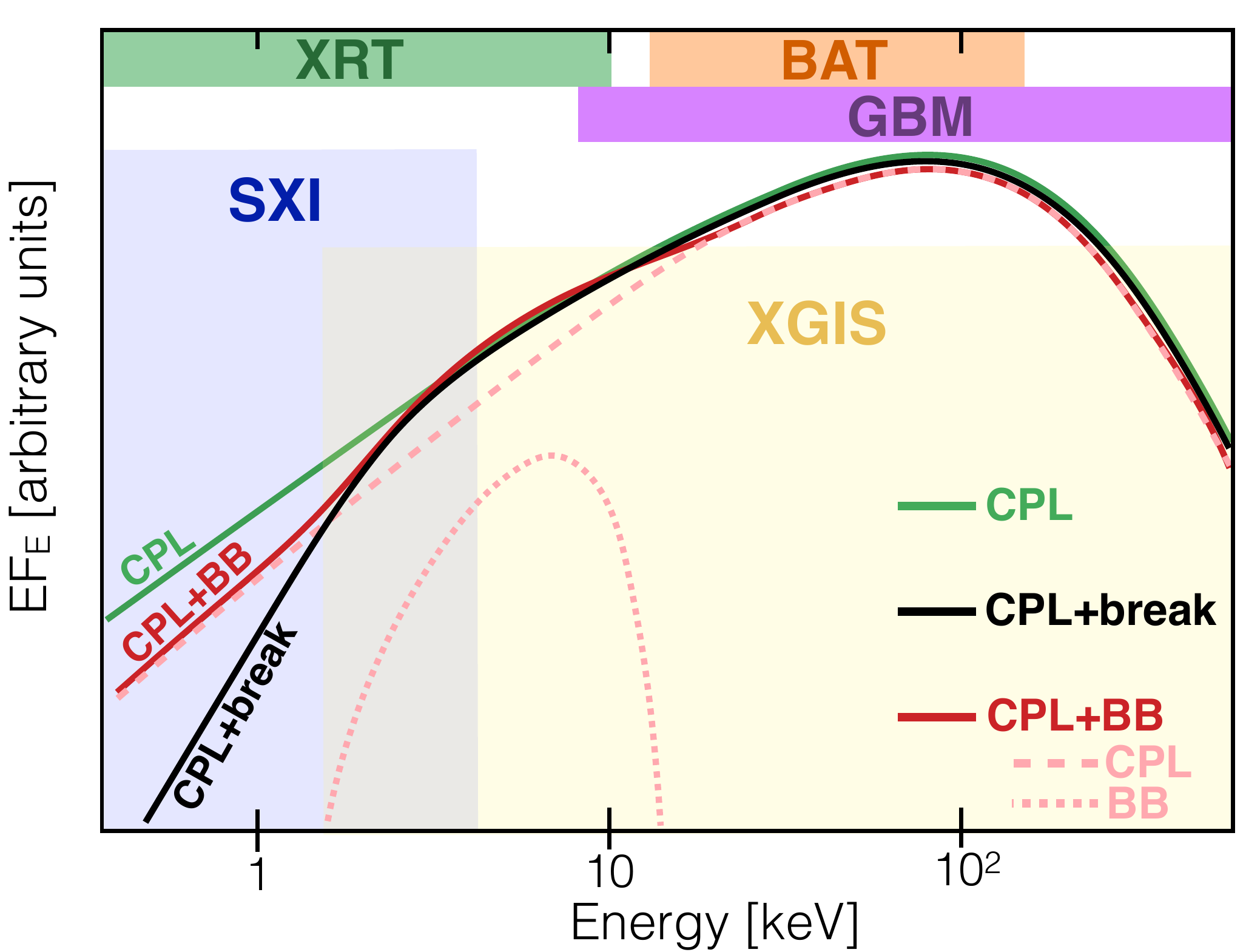}}
\caption{\footnotesize 
Left: distribution of the best fit values for the photon indices \aone\ and \atwo\ obtained from the time-resolved spectral analysis of 14 {\it Swift} GRBs with prompt emission detected simultaneously by XRT and BAT. \aone\ and \atwo\ are  the photon indices of the CPL+break model (black curve in the right-hand panel), below and above the break energy, respectively.  Dashed vertical lines mark the values expected for fast cooling synchrotron spectra. Adapted from \cite{oganesyan17}.
Right: schematic comparison between three different models: a standard a CPL (solid green), a CPL with a low-energy break (black solid line), and a CPL+BB (red solid line, the separated components are in dashed and dotted lines).
The three models have a very similar shape above a few keV, but they predict different behaviours at lower energies.
Shaded areas show the comparison between the sensitivity ranges of different instruments.}
\label{fig:sketch_alpha}
\end{figure*}

\section{The state-of-the-art}\label{sec:state-of-the-art}

At the present stage, prompt studies have been carried out mainly thanks to observations by BATSE ($\sim$\,2700 GRBs observed $>$\,25\,keV), {\it Fermi}-GBM ($\sim$\,2250 GRBs detected $>$\,8\,keV), and {\it Swift}-BAT ($\sim$\,1200 GRBs, $>$\,15\,keV). 
Spectral analyses of emission detected by these instruments showed that a smoothly connected broken PL is in most cases a good fit to the data. 
The spectral indices below and above the spectral peak have distributions centered around $\langle\alpha\rangle=-1$ and $\langle\beta\rangle=-2.3$. The position of the spectral peak \ep\ ranges between several keV and a few MeV, with typical values around 200\,keV.

In order to better characterise the prompt spectrum at low energy, where the inconsistency with synchrotron radiation is evident, \cite{oganesyan17} (O17 hereafter) considered GRBs with prompt emission (or part of it) detected simultaneously by XRT and BAT and performed joint spectral analysis.
These are mostly cases where BAT was triggered by a precursor and/or the prompt emission was particularly long and/or the repointing time particularly fast ($\lesssim60$\,s).

O17 collected a sample of 14 {\it Swift} GRBs with bright prompt emission simultaneously detected by both instruments, and performed time-resolved and time-integrated spectral analysis. 
In a later work, \cite{oganesyan18} (O18 hereafter) enlarged the sample to include additional twenty, fainter cases, for which only time-integrated analysis was possible.
Observations by the GBM where also included in the spectral analysis, when available (13 out of 34 events).

In both papers the same results have been found: in around 65\% of the analysed spectra, the XRT data lie below the low-energy PL extrapolation of the spectral shape outlined by BAT and GBM data. 
A spectral break is required to properly fit the entire spectrum. 
Note that the presence of the break is claimed only if, according to the $F$-test, the fit improves by more than 3$\sigma$ as compared to the fit provided by a model with no break. 
Below the break energy \eb, the spectrum is well described by a PL.
The overall spectrum can then be described by a function including: a low-energy PL \aone, a break energy \eb, a second PL \atwo, the $\nu F_\nu$ peak energy \ep, and eventually a third PL segment \be\ at high-energy.
The high-energy index \be\ is constrained only in few spectra, depending on the availability of GBM data, on the GRB brightness and on the location of \ep. 
In all the other cases, the high-energy part of the spectrum is satisfactorily modeled by an exponential cutoff. In most cases the best fit model is then a cutoff PL (CPL) with a low-energy break.
An example of this model is shown in Fig.~\ref{fig:sketch_alpha} (right-hand panel, black solid line named 'CPL+break').

The best fit values of the model parameters for the sample analysed in O18 are $\langle\alpha_1\rangle=-0.51$ ($\sigma=0.29$, for a gaussian fit to the distribution), $\langle\alpha_2\rangle=-1.54$ ($\sigma=0.26$), break energy \eb\ in the range 2-20\,keV, and \ep\ between 10\,keV and 1\,MeV. 
Fig.~\ref{fig:sketch_alpha} (left-hand panel) shows the distribution of the best fit values of $\alpha_1$ and $\alpha_2$ as inferred from the time-resolved analysis of the 14 brightest GRBs with prompt XRT observations (adapted from O17). Dashed vertical lines denote the synchrotron predicted values $\alpha^{\rm syn}_1=-2/3$ and $\alpha^{\rm syn}_2=-3/2$.

The joint analysis of  XRT+BAT(+GBM) spectral data is not a straightforward task, due to the necessity to model the effect of Galactic and intrinsic dust absorption, pile-up of photons in the XRT detector, and the uncertainties in the inter-calibration of the different instruments. 
The careful analysis and tests performed in O17 showed that the results on the presence of a feature at low energies are robust. 
Moreover, these results have been further strengthen by the discovery of a similar spectral shape in GRB~160625B, one of the brightest GBM bursts ever detected \citep{ravasio18}. 
A clear spectral break was constrained at around 100\,keV.
For this GRB, the analysis is performed using only the GBM instrument, that means above 8\,keV, where the emission is not affected by the unknown value of the intrinsic $N_{\rm H}$.
The time-resolved analysis shows no evolution of \aone\ and \atwo, a moderate hard-to-soft evolution of \eb, and a typical hard-to-soft evolution of \ep.
\begin{figure*}[t]
\begin{center}
{\includegraphics[clip=true,scale=0.55]{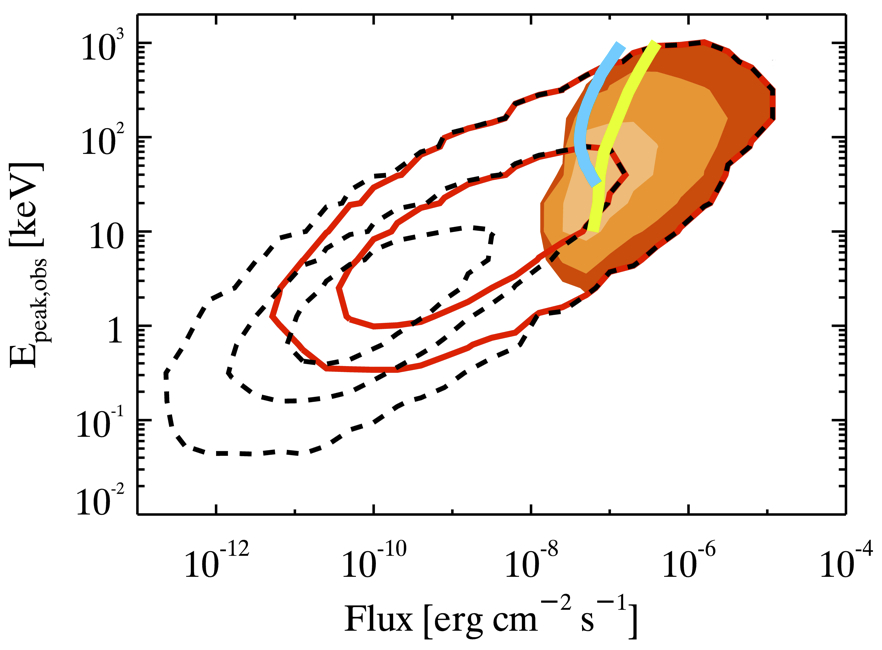}}
\caption{\footnotesize 
Contour plot (showing the 1,2,3\,$\sigma$ levels) for the population (adapted from \citealt{ghirlanda15}) of GRBs that will be detected by \th\ (red solid curves) in the $E_{\rm peak}$-flux plane (the flux is integrated in the 10-1000\,keV energy range). The subsample of events that will be detected by both SXI and XGIS for which a broad band spectral study will be possible is shown by the shaded contours. The solid yellow and cyan lines show the sensitivity limits of {\it Fermi} and BATSE, respectively, adapted from \cite{nava11}. The entire GRB population simulated in \cite{ghirlanda15} is shown by the dashed contour lines. }
\label{fig:epflux}
\end{center}
\end{figure*}

Some of the GRBs analysed in O17, O18 and in \cite{ravasio18} have been previously studied (in a non-systematic way) also by other authors. 
In all these studies, the analysis revealed the necessity for a more complex modeling than the standard Band or CPL models, in agreement with these more recent findings. 
However, in most cases the proposed modeling was very different and invoked the presence of two components: a non-thermal one (either Band, CPL or simple PL) and a BB-like spectrum.
The reason why the inclusion of a BB can account for the feature that in these recent studies is instead described as a spectral break is clear from Fig.~\ref{fig:sketch_alpha}, right-hand panel.
In this figure, a CPL with a low-energy break (black solid line) is compared with a CPL+BB model (red solid line).
As compared to the CPL function alone (dashed pink line), the break can appear as an excess of signal.
The inclusion of a BB (dotted pink curve) peaking (in $\nu F_\nu$) around the location of the break accounts for the apparent excess.
The resulting total model (CPL+BB, solid red line) is very similar to the case of a single component with a spectral break at low energy (black line).
For this reason, when both models are applied to the same spectral data, they usually return a similarly acceptable fit.
A statistical comparison between them with the aim of identifying the best model is not straightforward.
A comparison based on the reduced chi-square revealed that the reduced chi-square is systematically smaller when the break is considered, rather than when the BB component is added. 
In a few cases, O17, O18 and \cite{ravasio18} found that the break provides a statistically significant better fit. 
However, both models are in general acceptable, preventing us from reaching a firm conclusion.

The right-hand panel of Fig.~\ref{fig:sketch_alpha} shows that models can be ruled out if the available data extend well below the location of the break energy (or the location of the BB peak, in the two-component interpretation).

Summarising, the study of prompt spectra from 0.3\,keV was possible, thanks to XRT, only for 34 GRBs in 13 years of operations. 
Among these cases, around 20 have a clear feature at $\sim$\,keV energies. 
For a subsample of 4-5 GRBs, the feature is better described by a break rather than by the inclusion of a BB.
For the remaining cases, both models give an acceptable fit.

\th\ will allow the systematic study of spectra below 10\,keV, down to $\sim$\,0.3\,keV thanks to the SXI. The joint fit of SXI and XGIS will allow to model the spectrum over a large range of energies and discriminate between the two models.


\section{Beyond the state-of-the-art: THESEUS}\label{sec:theseus}
\begin{figure*}[t]
\begin{center}
{\includegraphics[clip=true,scale=0.23]{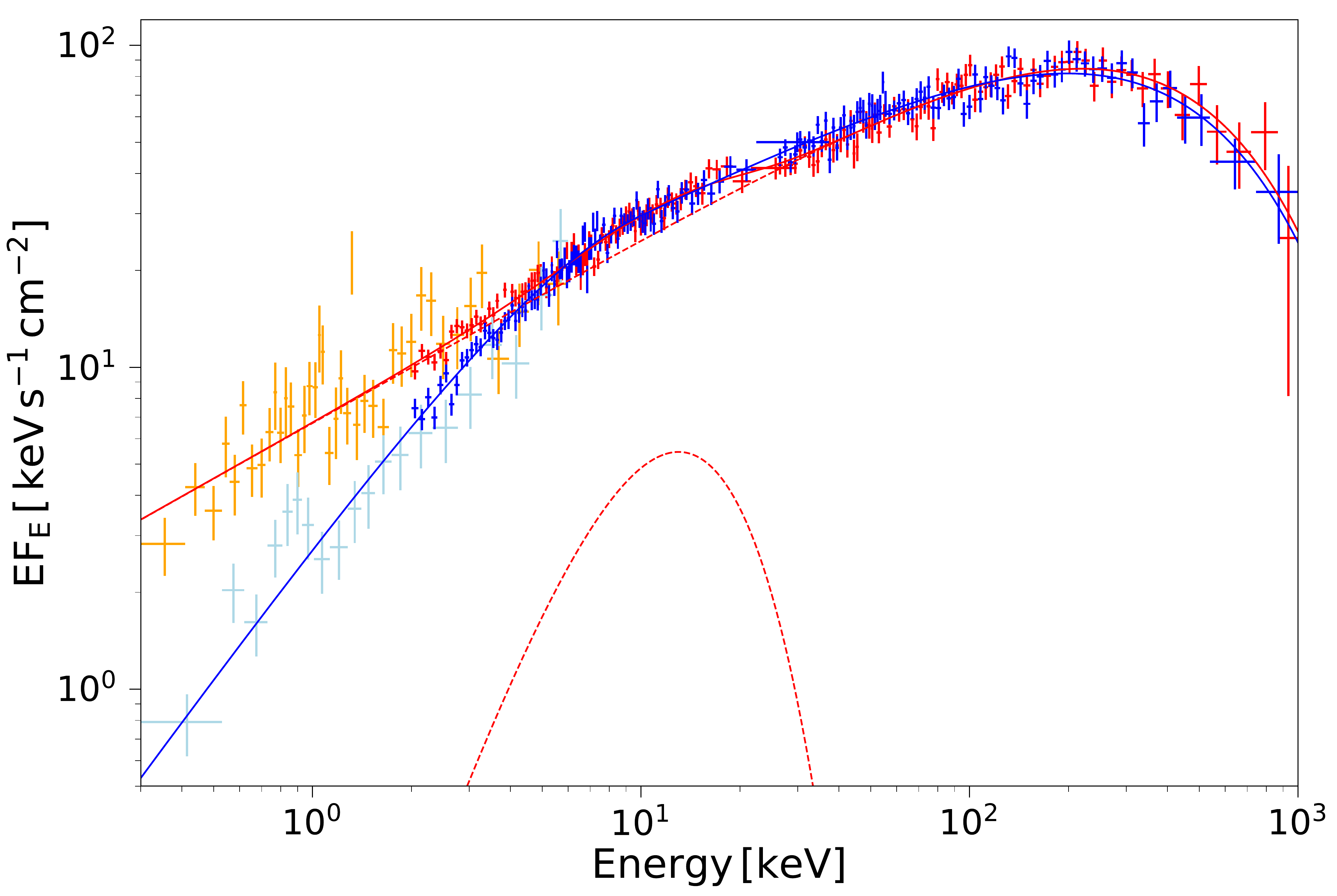}}
\caption{\footnotesize 
Simulation of a SXI and XGIS spectral data for two different models: CPL with a low-energy break (blue solid line) and a CPL+BB (red solid line for the sum of the two components, dashed red lines for the two separated components).
SXI data are in light-blue and orange. XGIS data are in blue and red.}
\label{fig:cpl+bb}
\end{center}
\end{figure*}

\th\ is a space mission concept proposed as M-class mission to the European Space Agency (ESA).
With the main focus of probing the early Universe, \th\ will be particularly suited for the study of X-ray transients, covering the energy range from 0.3\,keV to several MeV.
Sensitivity to emission in this energy range is achieved thanks to the Soft X-ray Imager (SXI, 0.3-6\,keV) and the X-Gamma ray Imaging Spectrometer (XGIS, 2\,keV-20\,MeV). For an exhaustive overview of instruments and scientific goals, the reader is referred to \cite{amati17} and \cite{stratta17}.

\th\ is expected to detect $\sim$\,400--800 GRBs per year. 
A sizable fraction will be simultaneously detected by SXI and XGIS during the prompt emission.
Fig.~\ref{fig:epflux} shows the density contours of the population of GRBs (simulated by \citealt{ghirlanda15} - dashed contours) that can be detected by \th\ (solid red lines). These bursts will allow us to explore the soft end of the GRB distribution in the $E_{\rm peak}$-Flux plane, now limited by the sensitivity of past and current detectors like {\it Fermi}-GBM and BATSE (solid yellow and cyan lines in Fig.~\ref{fig:epflux} - \citealt{nava11}).
In the lowest end of the detected population (solid red lines) will be located both low luminosity and high redshift events. Thanks to the combination of SXI and XGIS, \th\ will allow us to study the broad band (from 0.1\,keV to several MeV) spectrum of several of the detected events shown by the solid filled contours in Fig.~\ref{fig:epflux}.

To understand how a GRB prompt spectrum will be observed by \th\ and whether the detection of spectral breaks at $\sim$\,keV energies will be possible, we performed spectral simulations.
We simulated XGIS and SXI data, assuming three different models: i) CPL, ii) CPL with a break, and iii) CPL+BB.
For the second model, the photon indices below and above the break energy have been fixed to the synchrotron values, 
the peak energy has been fixed to 100\,keV, and the break energy \eb\ is around 10\,keV.
We chose an average flux (integrated between 0.1\,keV and 10\,MeV) equal to $5\times10^{-7}$\,erg\,cm$^{-2}$\,s$^{-1}$, and duration $T=20$\,s.
The chosen values for the column densities are $N_{\rm H,Galactic} = 5\times10^{20}$\, cm$^{-2}$, and $N_{\rm H,intrinsic} = 10^{22}$\,cm$^{-2}$, and the redshift is $z$= 2.

The remaining free parameters of the other two models (CPL and CPL+BB) have been chosen so that above the break energy, all three models have the same shape (as in the schematic example proposed in Fig.~\ref{fig:sketch_alpha}, right-hand panel), to reproduce the current observational picture. 
This resulted in a BB temperature $kT=2.85$\,keV and a total BB flux $F_{\rm BB}=8.1\times10^{-9}$erg\,cm$^{-2}$\,s$^{-1}$.
Current facilities would not be able to discriminate among the different models, unless the emission is detected with good statistics also by the XRT well below 10\,keV.
\begin{figure*}[t]
\begin{center}
{\includegraphics[clip=true,scale=0.23]{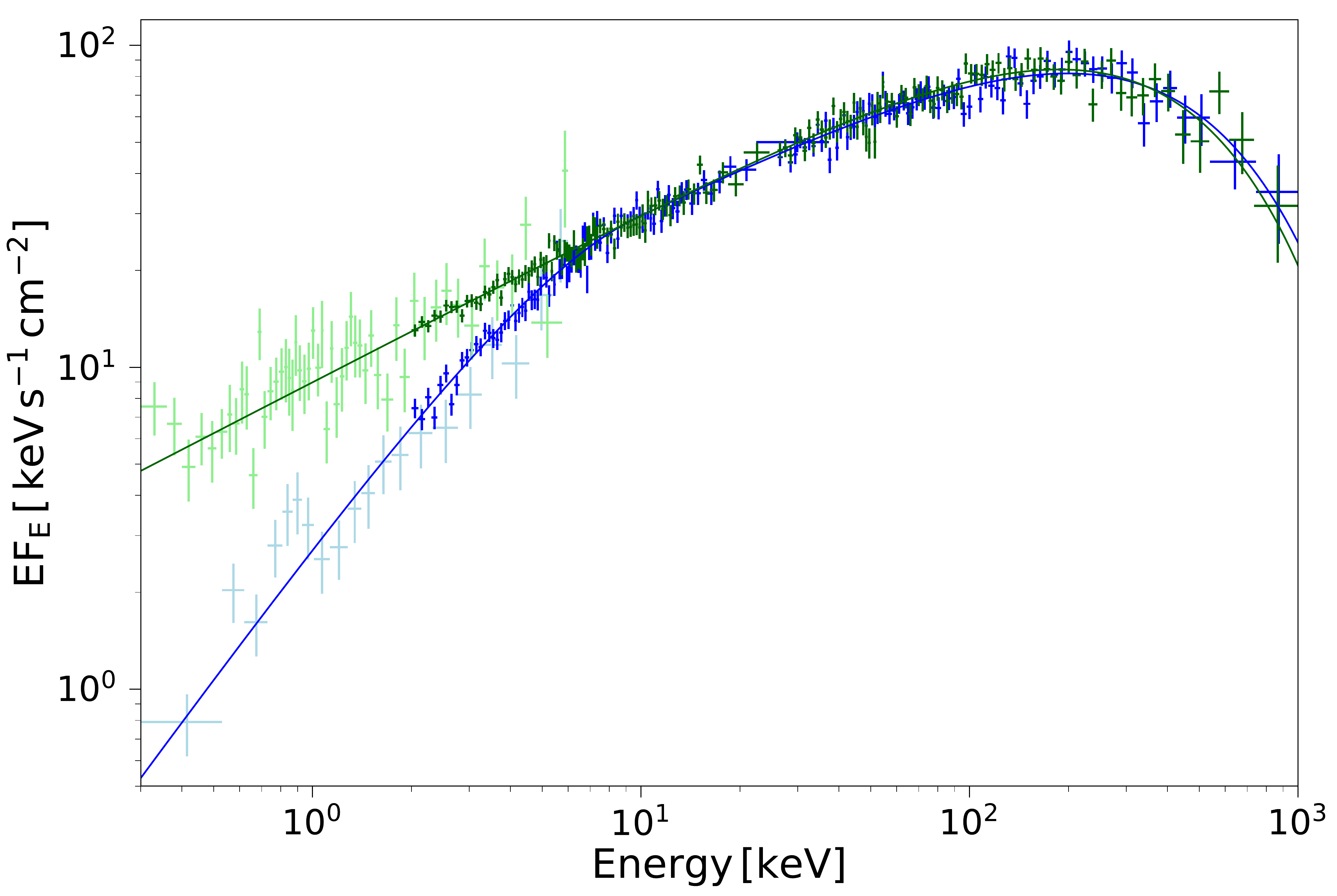}}
\caption{\footnotesize 
Simulation of a SXI and XGIS spectrum for two different models:  CPL with a low-energy break (blue solid line) and a CPL (green solid line).
SXI data are in light-green and light-blue. XGIS data are in green and blue.}
\label{fig:cpl}
\end{center}
\end{figure*}

For each model, we simulate the spectra as detected by \th\ adopting the following procedure.
We generate the fake SXI and XGIS spectra using the {\it fakeit} command in {\it XSPEC}. 
This procedure creates adjusted and randomized spectral files for the defined exposure time using instrumental responses and background files. 
For the SXI, only energy channels below 6\,keV are included.
For the low- and high-energy XGIS detectors, only the channels between
2 to 50\,keV and between 25\,keV to 1\,MeV (respectively) are considered. 
The energy channels are re-binned using the {\it grppha} tool, with the requirement of having at least 10 and 1000 counts in each channel for SXI and XGIS instruments, respectively. 

Once the simulated spectra have been obtained, 
a joint SXI+XGIS fit is performed, using Gaussian statistic.
The results of these simulations can be found in Fig.~\ref{fig:cpl+bb} and \ref{fig:cpl}. Note that in these figures we chose to show the de-absorbed best fit models and data.

In the first figure, (Fig.~\ref{fig:cpl+bb}), the CPL+break model (blue solid line) is compared to the CPL+BB model (red solid line).
The simulated XGIS data are shown in blue and red, and the SXI data are in light-blue and orange, respectively for the two models.
The simulated spectra are hardly distinguishable above $\sim$4\,keV, but they predict very different behaviours at lower energies. 
The difference between the two spectra is already visible in the low-energy channels of the XGIS instrument, and becomes evident with the inclusion of SXI data. 

Fig.~\ref{fig:cpl} reports the same CPL+break simulated spectrum (blue and light-blue) this time compared to a simple CPL case. 
The CPL model is shown by a green solid line. XGIS and SXI data are marked in dark-green and light-green, respectively.
Also in this case, the difference appears already clear below $\sim$4\,keV, in the lowest energy channels of the XGIS instrument. At even lower energies, the  flux predicted by the two models differs by a factor  2-to-10, much larger than the typical error on the SXI simulated spectral data.


\section{Conclusions}
\begin{figure*}[t]
\begin{center}
{\includegraphics[clip=true,scale=0.45]{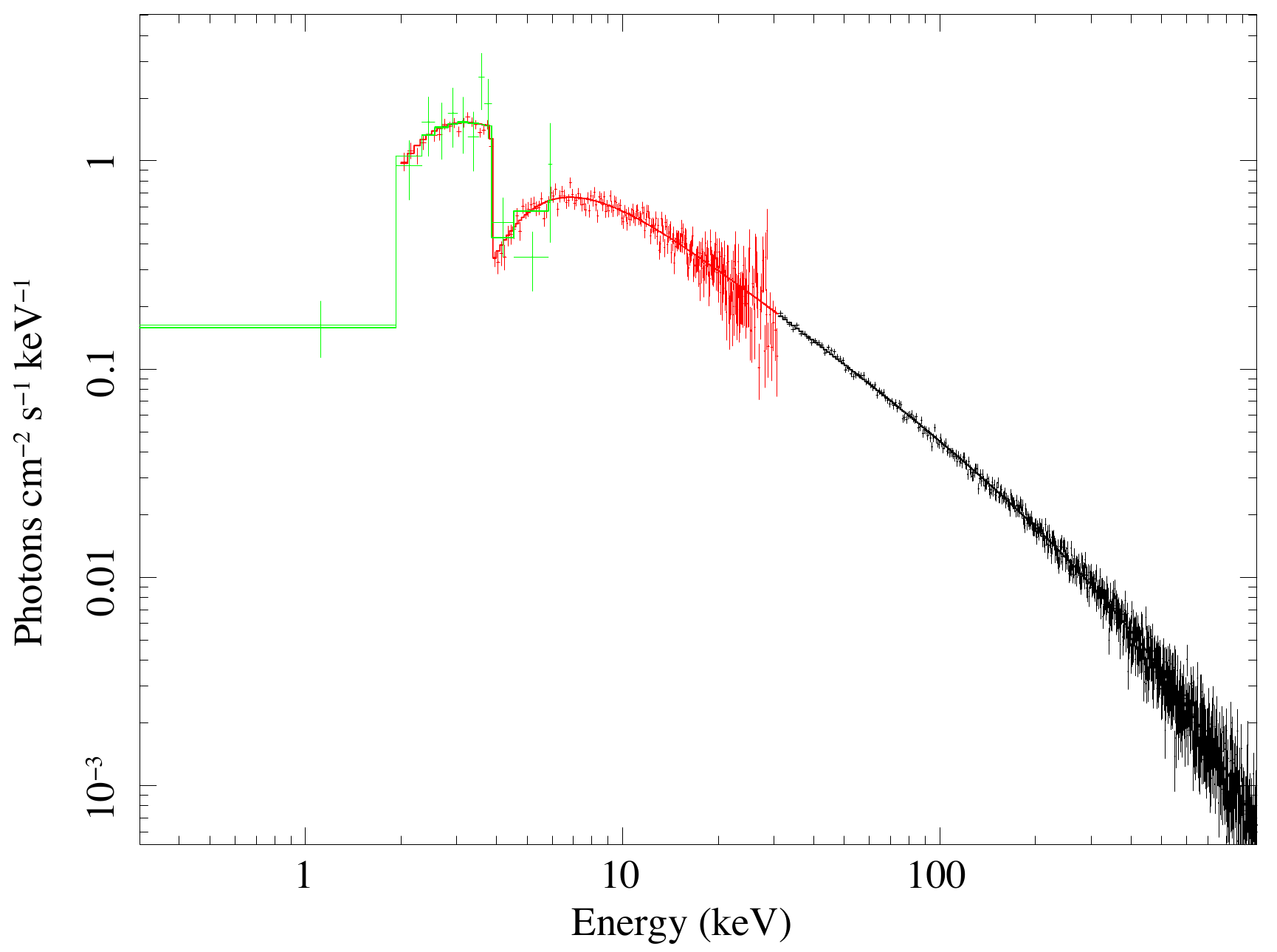}}
\caption{\footnotesize 
Simulation of the spectrum of the first 13\,s of the prompt emission of GRB\,990705 as would be measured by SXI (green) and XGIS (red and black). As can be seen, the transient absorption edge at $\sim$3.8\,keV and the high absorption column detected by the BeppoSAX/WFC \citep{amati00} would be revealed by {\it THESEUS} with very high significance.}
\label{fig:edge}
\end{center}
\end{figure*}

The long-lasting difficulties in the understanding of the origin of the prompt spectrum are strongly affecting our possibility of learning about the processes at work in GRB jets. 
The prompt spectra indeed carry the imprints of the properties of the emitting region (such as bulk Lorentz factor, magnetic field strength, particle spectra, distance from the central engine, jet composition), but the extraction of this information requires first a good understanding of the nature of the emission mechanism.

XRT observations have proved that a major breakthrough can definitely come from the study of prompt spectra below 10\,keV, an energy range rarely accessible with past and current facilities dedicated to prompt emission studies.
Recent analyses (O17 and O18) have shown that detailed modeling of the broad band (0.3\,keV -- few MeV) emission in bright long GRBs detected by {\it Swift} and {\it Fermi} revealed the common presence of an unexpected feature, typically between a few keV and 20\,keV.
This feature provides us with a new, powerful clue to finally identify the emission mechanism(s) at work.

These recent studies have shown that this feature can be satisfactorily modeled by including an additional PL segment at the low-energy end of the empirical model. Moreover, the best fit values of the PL indices are consistent with -2/3 and -3/2 below and above a characteristic break energy located between 2 and 20\,keV. 
Similar results have been found in one of the brightest events detected by {\it Fermi} \citep{ravasio18}, suggesting that the break energy has a distribution extending to large (but  uncommon) values.
On the other hand, in most cases, the observations can be accommodated also by a thermal+non--thermal model (e.g. CPL+BB).

The repointing timescale of Swift ($\sim$ 1 min on average) limits these studies to a very small number of cases (34 GRBs) for which the prompt emission lasted at least several tens of seconds or was preceded by a precursor event ($\sim$\,10\% of Swift bursts -- \citealt{burlon08}). 

Future observations by \th\ will unveil whether the feature at low-energy is ubiquitous, how it evolves with time, and whether a modeling in terms of a single (synchrotron) component is the correct one. We have indeed shown that the combined spectral analysis of XGIS and SXI will allow to discriminate between the synchrotron model and a thermal+non--thermal case.

We also remark that {\it THESEUS} will be able to detect and study with unprecedented accuracy possible absorption features in GRB X-ray prompt emission like the one detected by BeppoSAX in GRB\,990705 \citep{amati00} (Fig.~\ref{fig:edge}), which provide unique clues to the circum--burst environment and redshift determination.
 
\begin{acknowledgements}
L.N. acknowledges funding from the European Union's Horizon 2020 Research and Innovation programme under the Marie Sk\l odowska-Curie grant agreement n.\,664931.
\end{acknowledgements}

\bibliographystyle{aa}

\end{document}